%% file: PRL_Btautau.tex
\documentclass[twocolumn,showpacs,aps,prl,superscriptaddress]{revtex4}

\usepackage{graphicx}
\usepackage{dcolumn}
\usepackage{amsmath}
\usepackage{epsfig}

\input babarsym

\newcommand{\BABARPubYear}    {05}
\newcommand{\BABARPubNumber}  {36}
\newcommand{\SLACPubNumber} {11558}

\def\figurebox#1#2#3{%
    \def\arg{#3}%
    \ifx\arg\empty
    {\hfill\vbox{\hsize#2\hrule\hbox to #2{\vrule\hfill\vbox to #1{\hsize#2\vfill}\vrule}\hrule}\hfill}%
    \else
    {\hfill\epsfbox{#3}\hfill}%
    \fi}

\begin{document}

\preprint{\babar-PUB-\BABARPubYear/\BABARPubNumber} 
\preprint{SLAC-PUB-\SLACPubNumber} 

\begin{flushleft}
BABAR-PUB-\BABARPubYear/\BABARPubNumber\\
SLAC-PUB-\SLACPubNumber\\
\end{flushleft}

\title{
{\large \bf
A Search for the Rare Decay $\mathbf{B^{0} \rightarrow \tau^{+} \tau^{-}}$ at \babar} 
}

\input authorsl

\date{\today}

\begin{abstract}
\input{abstract}
\end{abstract}


\maketitle

\begin{figure}[t]
\begin{center}
\resizebox{1.3in}{!}{\includegraphics*[2.8in, 8.2in][4.1in,9.4in]{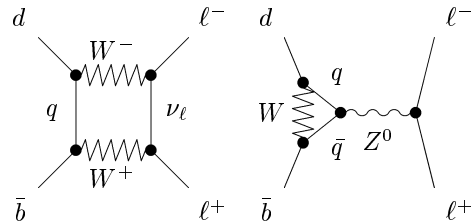}}
\resizebox{1.3in}{!}{\includegraphics*[5.4in, 8.2in][6.7in,9.4in]{box_penguin.ps}}
\caption{Standard model box and penguin processes that can mediate $B^0 \rightarrow \ell^+ \ell^-$ ($q=t,c,u$). 
\label{diagramFigure}}
\end{center}
\end{figure}

None of the leptonic decays $B^0 \rightarrow \ell^+ \ell^-$
($\ell=e,\mu,\tau$) have been observed. 
In the standard model of particle physics, the decays can be mediated
by box and penguin diagrams (Fig.\,\ref{diagramFigure}). 
The standard model produces only the combinations $\ell_{R}^+ \ell_{L}^-$ and $\ell_{L}^+ \ell_{R}^-$. The amplitudes for the decay of a spin-zero particle to these states are proportional to $m_{\ell}$ and thus the decay rates are suppressed by $(m_{\ell}/m_{B})^2$. The suppression is smallest for $B^{0} \rightarrow \tau^+ \tau^-$ due to the large $\tau$ mass. 
 The standard model prediction for the $B^{0} \rightarrow \tau^+ \tau^-$ branching fraction is \cite{Babar:1998}

\begin{eqnarray}
\mathcal{B}^{SM}(B^{0} \rightarrow \tau^{+}\tau^{-}) & = &  1.2 \times
10^{-7} 
\nonumber\\
& &  \times  \left[ \frac{f_{B}}{200\,\mbox{MeV} } \right]^2 \left[ \frac{\vert
V_{td} \vert}{0.007} \right]^2 ,
\end{eqnarray}

\noindent where $f_{B}$ is the $B$ decay constant and $V_{td}$ is the Cabibbo-Kobayashi-Maskawa matrix element. The theoretical uncertainty on $f_{B}$ and the experimental error on $V_{td}$ dominate the uncertainty on the predicted branching fraction.

Extensions of the standard model containing leptoquarks, which couple
leptons to quarks, predict enhancements for $\mathcal{B}(B^0
\rightarrow \tau^+ \tau^-)$ \cite{Grossman:1997} that are
proportional to the square of the leptoquark coupling. In theories
that contain two Higgs doublet fields, the rate can be enhanced by
powers of $\tan \beta$, the ratio of vacuum expectation values of the
two Higgs doublet fields \cite{Logan:2000, Babu:2000}. Since $B^0
\rightarrow \ell^+ \ell^-$ has not been observed, one can only
constrain model parameters using the measured branching
fraction limits. While $\tan \beta$ is constrained by all three modes
($\ell=e,\mu,\tau$), only $B^0 \rightarrow \tau^+
\tau^-$ can constrain the coupling of a leptoquark to the third
lepton generation or other new physics involving only the third
generation. 

The analysis described here provides the first upper limit on $\mathcal{B}(B^0 \rightarrow \tau^+ \tau^-)$. The data were collected with the \babar\ detector at the asymmetric PEP-II $e^{+}e^{-}$ storage ring. A full description of the \babar\ detector is given in Ref.\,\cite{babar:2002}. In brief, charged-particle momenta are measured with a tracking system comprising a silicon vertex detector (SVT) and a drift chamber (DCH) placed within a highly uniform $1.5$-T magnetic field generated by a superconducting solenoid. Electron and photon energies are measured with an electromagnetic calorimeter (EMC) constructed with Thalium-doped CsI scintillating crystals. Muons are distinguished from hadrons in a steel magnetic-flux return instrumented with resistive plate chambers (IFR). Charged particle identification is provided by a Cherenkov detector (DIRC) and the tracking system.
The data sample consists of $210\,$fb$^{-1}$ collected at the peak of the $\Upsilon(4S)$ resonance, which corresponds to $232 \pm 3$ million $B \bar{B}$ pairs. The expected background and the expected signal efficiency are obtained from Monte Carlo simulation samples. The sample events were generated with the EvtGen event simulator \cite{Lange:2001} and propagated through a detailed model of the \babar\ detector using the GEANT4 detector simulator \cite{Geant4:2003}. 

Isolating $B^{0} \rightarrow \tau^{+}\tau^{-}$ poses a unique challenge. This decay contains at least two and as many as four neutrinos, so there is no kinematic discriminant that separates signal from background due to undetected particles. Since two $B$ mesons are produced in an $\Upsilon(4S)$ decay, the misassignment of decay products to the parent $B$ must be avoided. 
We completely reconstruct one $B$ candidate in each event (hereafter referred to as the companion $B$) and search for the signal decay among the remaining detected particles. The combinatorial background in the companion-$B$ reconstruction is determined by a fit to the companion-$B$ invariant mass distribution.
We employ the parameters

\begin{eqnarray}
m_{\mbox{\tiny ES}} & = & \sqrt{E_{beam}^{* 2} - p_{B}^{* 2}} \\ 
\Delta E & = & E_{B}^{*}-E_{beam}^{*},
\end{eqnarray}

\noindent where $p_{B}^{*}$ and $E_{B}^{*}$ are the reconstructed companion-$B$ momentum and energy in the center-of-mass (CM) frame. $E_{beam}^{*}$ is the beam energy in the CM frame. The $m_{\mbox{\tiny ES}}$ distributions are fit with a probability density function composed of a Crystal Ball function \cite{crystalnote} to model the peak at the $B$ mass and an ARGUS function \cite{argusnote} to model the nonpeaking combinatorial background.

The companion $B$ is fully reconstructed in a hadronic mode $\bar{B}^{0} \rightarrow D^{(*)}X$ , where $D^{(*)}$ is either a $D^{+}$ \cite{chargenote}, $D^0$, or $D^{* +}$ and $X$ is a system consisting of up to five particles of the type $\pi^{\pm},\pi^{0},K^{\pm}$, or $K_{S}^0$ \cite{vub:2004}. $D^{* +}$ mesons are reconstructed in the channel $D^{0} \pi^{+}$. $D^0$ mesons are reconstructed in the channels $K^- \pi^+$, $K^- \pi^+ \pi^0$, $K^- \pi^+ \pi^- \pi^+$, and $K_{S}^0 \pi^+ \pi^-$. $D^+$ mesons are reconstructed in the channels $K_{S}^0 \pi^+$, $K^- \pi^+ \pi^+$, $K_{S}^0 \pi^+ \pi^0$, $K_{S}^0 \pi^+ \pi^+ \pi^-$, and $K^+ K^- \pi^+$. The $\Delta E$ of the companion $B$ is required to be within two mode-dependent standard deviations of the mean when no $\pi^0$ is present, or to satisfy $-0.09 < \Delta E < 0.06$\,GeV for reconstructions with one or more $\pi^0$. If more than one $B$ candidate is reconstructed in the same mode, the reconstructed $B$ with the smallest $\vert \Delta E \vert$ is selected. 
For each mode, the purity $B_{pur}$ is the ratio of the number of
events before signal selection in the fitted peak to the total number
of events in the region $5.27<m_{\mbox{\tiny ES}} < 5.29$\,GeV. Only
events reconstructed in a mode with $B_{pur}>0.12$ are selected, which
results in the reconstruction of $147$ distinct modes in the data
sample. If $B$ candidates are reconstructed in more than one mode, the $B$ reconstructed in the mode with the highest $B_{pur}$ is selected as the companion $B$. 

\begin{figure}[tbp]
\begin{center}
\resizebox{3.45in}{!}{\includegraphics{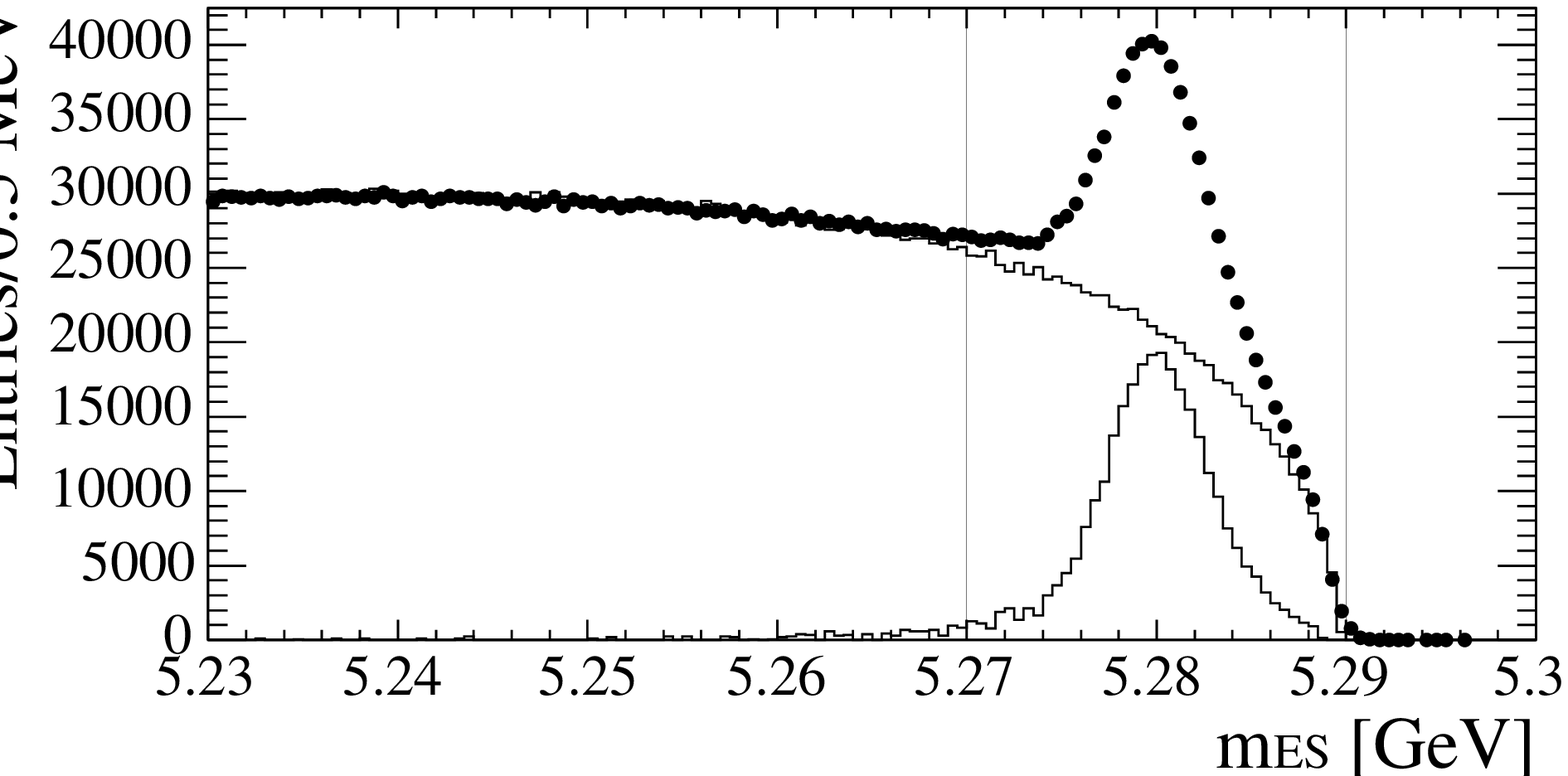}}
\resizebox{3.45in}{!}{\includegraphics{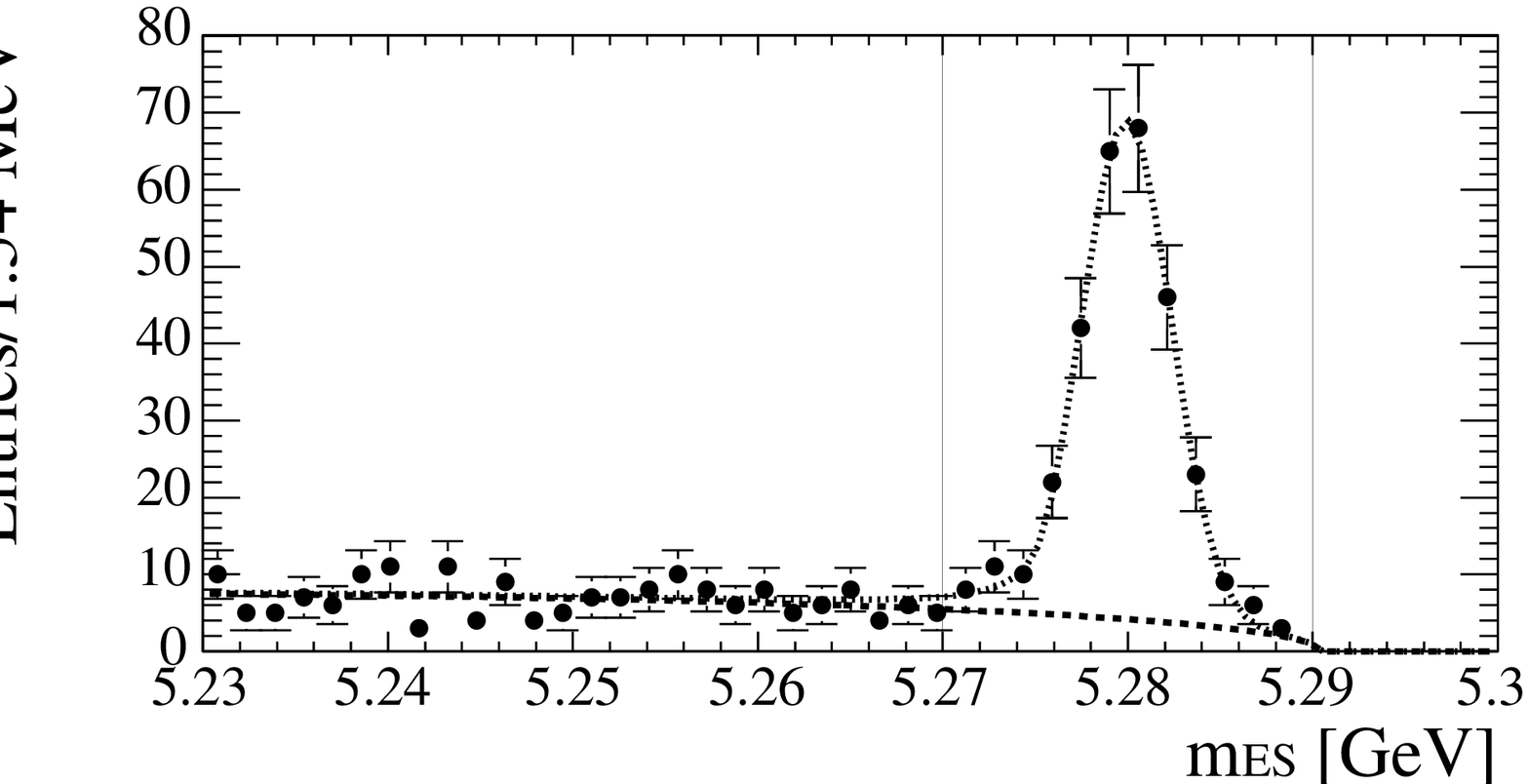}}
\caption{Above, the $m_{\mbox{\tiny ES}}$ distribution for the
  hadronic companion $B$ in data (dots) and scaled simulated background
  (upper histogram) before the signal $B^0 \rightarrow \tau^+ \tau^-$
  selection is applied; the lower histogram is obtained by subtracting
  the background from the data. The companion-$B$ yield is $N_{B^0
  \bar{B}^0}=(2.80 \pm 0.27) \times 10^{5}$. Below, the $m_{\mbox{\tiny ES}}$
  distribution after the signal $B^0 \rightarrow \tau^+ \tau^-$
  selection. The fitted probability density function
  (short-dash) and its ARGUS component (dash) are superimposed on the
  data (dots). We obtain $N_{obs}=263 \pm 19$ events in the peak. \label{mesFigure}}
\end{center}
\end{figure}

We estimate the total companion-$B$ yield from all reconstructed modes
using the $B \bar{B}$ and $q \bar{q}$ ($q=u,d,s,c$) simulated samples
before applying the signal $B^0 \rightarrow \tau^+ \tau^-$
selection. We first remove the peak from the $B^0 \bar{B}^0$ simulated
sample using the fitted Crystal Ball probability density function. Subtracting the simulated combinatorial background $m_{\mbox{\tiny ES}}$ shape, fitted to the data below $5.26$\,GeV, from the data distribution yields a nominal companion-$B$ yield of $N_{B^0 \bar{B}^0}=(2.80 \pm 0.27) \times 10^{5}$ (Fig.\,\ref{mesFigure}). The systematic error on $N_{B^0 \bar{B}^0}$ is estimated to be $10 \%$ by varying the fit region and by varying the combinatorial background composition with event-shape-variable cuts.

The companion-$B$ decay products are removed from the event and the signal-$B$ characteristics are sought among the remaining particles. The dominant background to $B^0 \rightarrow \tau^+ \tau^-$ arises from decays $b \rightarrow W^{-} c(\rightarrow W^{+}s)$, in which the $s$ quark hadronizes into a $K_{L}^0$ that escapes detection and the virtual $W^+$ and $ W^{-}$ mimic the virtual $W^+$ and $ W^-$ emitted by the signal $\tau$. A secondary background originates in events in which two oppositely charged particles are lost outside the detector fiducial region.
We select signal events that are consistent with each $\tau$ decaying to a single charged particle (and one or two $\nu$) by selecting events with zero net charge and two tracks in the recoil system. Each track must leave at least twelve hits in the DCH, originate within $10$ cm of the beamspot in the beam direction and within $1.5$ cm in the transverse direction, and have a transverse momentum of at least $0.1$\,GeV. To eliminate background originating from $b \rightarrow W^{-} c(\rightarrow W^{+}s)$ events, the selection rejects events with identified $K^{+}$, $K_{S}^0$, or $K_{L}^0$. The $K^{\pm}$ candidates are identified by a neural network with inputs taken from the SVT, the DCH, and the DIRC. The $K_{S}^0$ candidates are identified as a $\pi^+\pi^-$ pair with invariant mass consistent with the $K_{S}^0$ mass ($0.473<m_{\pi^{+} \pi^{-}}<0.523$\,GeV). The $K_{L}^0$ candidates are identified from clusters in the EMC that have not been associated with a charged track or included in a candidate $\pi^0$. A neural network is employed to identify $K_{L}^0$ candidates using the cluster energy and shower-shape variables, which discriminate hadronic from electromagnetic showers.

The multiplicities of $e$, $\mu$, and $\pi^{0}$ in the recoil system must be consistent with each $\tau$ decaying in one of the channels $\tau \rightarrow \pi \nu, \rho \nu$, $e \nu \bar{\nu}$ or $\mu \nu \bar{\nu}$ (Table \ref{signalTable}). The $e$ candidates are identified with $dE/dx$ measurements from the DCH and shower shape variables from the EMC. The $\mu$ candidates are identified with variables from the IFR (to reject the $\pi$ hypothesis) and EMC (to reject the $e$ hypothesis). Track candidates that are not identified as $e$, $\mu$ or $K$ are assumed to be $\pi$. Events with $\pi^0$ are vetoed unless the $\pi^0$ can be associated to a $\pi^+$ such that the invariant mass is consistent with the $\rho$ mass ($0.6<m_{\pi^{+}\pi^{0}}<1.0$\,GeV). The $\pi^0$ candidates are formed from pairs of $\gamma$ candidates with invariant mass $0.090<m_{\gamma \gamma}<0.170$\,GeV, with each $\gamma$ having an energy greater than $0.030$\,GeV. Since the presence of residual unassociated energy in the EMC ($E_{res}$) is a strong indication that an unreconstructed $\pi^0$ or $K^0$ is present, we require $E_{res}<0.11$\,GeV. 

 \begin{table}[tbp]
 \begin{center}
 \caption{Signal $B^0 \rightarrow \tau^+ \tau^-$ branching fraction and requirements by mode ($\ell = e, \mu$). \label{signalTable}}
\vspace{2mm}
 \begin{tabular}{l c c c c c} \hline \hline
  Selection Mode & $\mathcal{B}(\%)$ \cite{Eidelman:2004} & $N_{e}+N_{\mu}$ & $N_{\pi^{0}}$ & $m_{\pi \pi^{0}}$  \\ \hline 
 $\tau^+ \tau^- \rightarrow \ell \nu \bar{\nu}/\ell^{\prime} \nu \bar{\nu}$ & $12.4$ & $2$ & $0$ &  NA  \\
 $\tau^+ \tau^- \rightarrow \ell \nu \bar{\nu}/ \pi \nu  $ & $7.8$ & $1$ & $0$ & NA  \\ 
 $\tau^+ \tau^- \rightarrow \ell \nu \bar{\nu}/\rho \nu  $ & $17.7$ & $1$ & $1$ & $[0.6,1.0]$\,GeV \\
 $\tau^+ \tau^- \rightarrow \pi \nu/\pi \nu     $ & $1.2$ & $0$ & $0$ & NA  \\
 $\tau^+ \tau^- \rightarrow \pi \nu/ \rho \nu   $ & $5.6$ & $0$ & $1$ & $[0.6,1.0]$\,GeV \\
 $\tau^+ \tau^- \rightarrow \rho \nu / \rho \nu $ & $6.3$ & $0$ & $2$ & $[0.6,1.0]$\,GeV \\ \hline \hline
 \end{tabular}
 \end{center}
 \end{table}

\begin{table}[tbp]
\begin{center}
\caption{$\epsilon_{sig}$, $N_{expected}$  and $N_{obs}$ obtained from
  individual fits by signal mode. The errors are statistical and fit
  error added in quadrature. Branching fractions are included in the
  efficiency estimates. The $\pi \nu / \pi \nu$ channel is dominated 
  by crossfeed from other signal channels. \label{bymodeTable}}
\vspace{2mm}
\begin{tabular}{l c c c } \hline \hline
Selection Mode & $\epsilon_{sig} (\%)$  & $N_{expected}$  & $N_{obs}$ \\ \hline
$\tau^+ \tau^- \rightarrow  \ell \nu \bar{\nu}/\ell^{\prime} \nu \bar{\nu}$ & $0.9 \pm 0.2 $ & $46 \pm 4$ & $54 \pm 7$\\
$\tau^+ \tau^- \rightarrow \ell \nu \bar{\nu}/ \pi \nu$ &  $1.5 \pm 0.3$ & $122 \pm 6$  & $105 \pm 11$\\
$\tau^+ \tau^- \rightarrow \pi \nu/\pi \nu$  & $1.5 \pm 0.3$ & $89 \pm 6 $  &  $80 \pm 11$\\
$\tau^+ \tau^- \rightarrow \rho \nu / \rho \nu$  & $0.3 \pm 0.1$ & $21 \pm 3 $ & $ 15 \pm 6$ \\ \hline \hline
\end{tabular}
\end{center}
\end{table}

The $\tau$-daughter candidates are Lorentz-boosted with the companion-$B$ momentum.
While distributions of the momenta $\mathbf{p}_+$ and $\mathbf{p}_-$ of the charged daughters exhibit no discrimination from the background momentum distributions, correlations among $\vert \mathbf{p}_+ \vert $, $\vert \mathbf{p}_- \vert$, and $\cos \theta \equiv \mathbf{p}_+ \cdot \mathbf{p}_- / \vert \mathbf{p}_+ \vert \vert \mathbf{p}_- \vert$ afford some discrimination, especially when categorized by signal $B^0 \rightarrow \tau^+ \tau^-$ selection mode. Cascade decay background events manifest an asymmetry in  $\vert \mathbf{p}_+ \vert $ and $\vert \mathbf{p}_- \vert$ that is not present in signal events. The parameters $\vert \mathbf{p}_+ \vert$, $\vert \mathbf{p}_- \vert$, $\cos \theta$, $E_{res}$, and the selection mode are used as inputs in a neural-network analysis trained to discriminate signal from background. The final selection requirement is a neural network output ($NN$) consistent with signal events. 

The signal $B^0 \rightarrow \tau^+ \tau^-$ selection criteria for $E_{res}$, $NN$ and $B_{pur}$ are chosen to minimize the expected upper limit on $\mathcal{B}(B^0 \rightarrow \tau^+ \tau^-)$. That optimization also rejects the signal selection modes $\tau^+ \tau^- \rightarrow \ell \nu \bar{\nu}/\rho \nu$ and $\tau^+ \tau^- \rightarrow \pi \nu/ \rho \nu $.
After the full signal $B^0 \rightarrow \tau^+ \tau^-$ selection, the combinatorial companion-$B$  background is estimated and subtracted using ARGUS and Crystal Ball fits to the $m_{\mbox{\tiny ES}}$ distributions in simulation samples and data (Fig.\,\ref{mesFigure}). From these fits we determine the signal efficiency ($\epsilon_{sig}$), the expected number of background events ($N_{expected}$), and the number of observed data events ($N_{obs}$). Including systematic uncertainties described below, we obtain $\epsilon_{sig}  =  0.043 \pm 0.009$, and $N_{expected} = 281 \pm 48$. We extract from the fit $N_{obs}=263 \pm 19$ events in the data after the full selection. The central value of the $B^0 \rightarrow \tau^+ \tau^-$ branching fraction is $(-1.5 \pm 4.4) \times 10^{-3}$. We find no evidence for signal events. Table \ref{bymodeTable} shows $\epsilon_{sig}$, $N_{expected}$ and $N_{obs}$ obtained from individual fits to specific signal selection modes.

Systematic uncertainties on $N_{expected}$ and $\epsilon_{sig}$ arise from several sources. The simulation statistical uncertainty for $N_{expected}$ ($\epsilon_{sig}$) is $10$ events ($11\%$). 
The systematic uncertainties are estimated for cluster corrections to be $8$ ($3\%$), for particle identification corrections $10$ ($10 \%$), and for tracking corrections $7$ ($3 \%$).
The $m_{\mbox{\tiny ES}}$ background subtraction fits after the full selection add a further systematic uncertainty of $4$ ($2\%$). 
We estimate the systematic uncertainty on $N_{expected}$ due to $B$ decay modeling in
EvtGen to be $10\%$. We estimate the systematic uncertainty due to our model of $\tau$ decay by inserting distributions obtained from the specialized $\tau$ Monte Carlo TAUOLA \cite{Golonka:2003} to decay two $\tau$ produced with the same helicity and the requisite momentum for a $B^0 \rightarrow \tau^+ \tau^-$ decay.
For each simulated event, the decay mode of each $\tau$ is identified and the $\vert \mathbf{p}_+ \vert$, $\vert \mathbf{p}_- \vert$ and $\cos \theta$ values are replaced with values sampled from distributions generated by TAUOLA for that mode.
The relative $\epsilon_{sig}$ variation between EvtGen and TAUOLA simulation is $2 \%$.

\begin{figure}[tbp]
\begin{center}
\resizebox{3.5in}{!}{\includegraphics{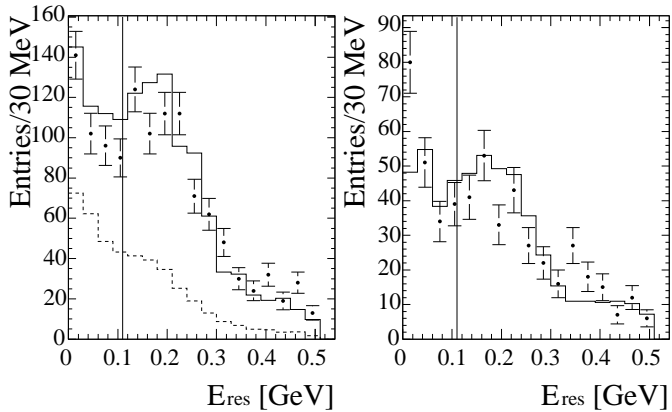}}
\caption{The $E_{res}$ distribution in the nominal sample (left) and
  the control sample (right) for data (dots), simulated background
  (solid histogram), and simulated signal (dashed histogram). The
  simulated signal distribution normalization is arbitrary. All
  requirements except those for $E_{res}$ and NN are imposed. The
  events to the left of the vertical line are selected.
  \label{eresFigure}}
\end{center}
\end{figure}

A final systematic uncertainty for both signal and background is
assigned to the modeling of $E_{res}$. The simulation of background
hits and hadronic interactions in the EMC does not perfectly model the
data, and the discrepancy manifests itself in the $E_{res}$
distribution (Fig.\,\ref{eresFigure}). This uncertainty is estimated from the difference between data and the simulation for a control process.
The control sample selection is identical to the $B^0 \rightarrow \tau^+ \tau^-$ selection except that events with an additional reconstructed $K_S^0$ are selected and the $K_S^0$ daughters are removed from the event. For correct $K_{S}^0$ reconstructions, this control sample models the $K_{L}^0$ background while for $K_{S}^0$ reconstructions from random combinations of tracks it models the 
backgrounds in which two oppositely charged particles are lost due to the limited detector acceptance in the direction of the higher energy beam.
The composition of the background in the simulated control sample agrees well with that of the simulated signal sample.
The control sample yields are $135 \pm 14$ events (data) and $125 \pm 7$ (simulation), for a relative discrepancy of $(8 \pm 13) \%$, consistent with zero. The systematic uncertainty due to modeling the residual energy in the EMC is taken to be the uncertainties in data and simulation yields added in quadrature, namely $13 \%$.

Systematic uncertainties  on the companion-$B$ yield, expected background, and $\epsilon_{sig}$ are folded into the upper limit calculation using the technique described in Ref. \cite{Barlow:2002}, giving 

\begin{eqnarray}
\mathcal{B}(B^{0} \rightarrow \tau^{+}\tau^{-}) & < & 4.1 \times 10^{-3} ,
\end{eqnarray}

\noindent at the $90 \%$ confidence level. 
The result constrains leptoquark couplings as described in Ref.\,\cite{Grossman:1997}. For example, the scalar $SU(2)$ doublet leptoquark $S_{1/2}$ can mediate $B^0 \rightarrow \tau^{+}\tau^{-}$. If no other leptoquark mediates the decay, the product of its coupling $\lambda_{R}^{33}$ (coupling righthanded $b$ with $\tau$) with $\lambda_{R}^{13}$ (coupling righthanded $d$ with $\tau$) is

\begin{eqnarray}
\lambda_{R}^{33} \lambda_{R}^{13} & < & 1.4 \times 10^{-2} \left[ \frac{m_{S_{1/2}}}{100 \, \mbox{GeV}} \right]^2,
\end{eqnarray}

\noindent at the $90 \%$ confidence level, where $m_{S_{1/2}}$ is the $S_{1/2}$ mass.

\input acknow_PRL.tex

\bibliography{PRL_Btautau}

\end{document}

%% file: authorsl.tex
%% author list as of 28-Jun-2005 (635 authors)
%
\author{B.~Aubert}
\author{R.~Barate}
\author{D.~Boutigny}
\author{F.~Couderc}
\author{Y.~Karyotakis}
\author{J.~P.~Lees}
\author{V.~Poireau}
\author{V.~Tisserand}
\author{A.~Zghiche}
\affiliation{Laboratoire de Physique des Particules, F-74941 Annecy-le-Vieux, France }
\author{E.~Grauges}
\affiliation{Universit\"at de Barcelona (IFAE) Fac.\ Fisica.\ Dept.\ ECM Avda Diagonal 647, 6a planta E-08028 Barcelona, Spain }
% ...
\collaboration{The \babar\ Collaboration}
\noaffiliation

%% file: abstract.tex
We present the results of a search for the decay $B^0 \rightarrow \tau^+\tau^-$
in a data sample of $(232 \pm 3) \times 10^6$ $\Upsilon(4S) \rightarrow
B \bar{B}$ decays using the \babar\ detector. Certain extensions of the Standard Model predict measurable levels of this otherwise rare decay. 
We reconstruct fully one neutral $B$ meson and seek evidence for the signal decay in the rest of the event.
We find no evidence for signal events and obtain $\mathcal{B}(B^0 \rightarrow \tau^+\tau^-) < 4.1 \times 10^{-3}$ at the $90 \%$ confidence level.

%% file: acknow_PRL.tex
We are grateful for the excellent luminosity and machine conditions
provided by our \pep2\ colleagues, 
and for the substantial dedicated effort from
the computing organizations that support \babar.
The collaborating institutions wish to thank 
SLAC for its support and kind hospitality. 
This work is supported by
DOE
and NSF (USA),
NSERC (Canada),
IHEP (China),
CEA and
CNRS-IN2P3
(France),
BMBF and DFG
(Germany),
INFN (Italy),
FOM (The Netherlands),
NFR (Norway),
MIST (Russia), and
PPARC (United Kingdom). 
Individuals have received support from CONACyT (Mexico), A.~P.~Sloan Foundation, 
Research Corporation,
and Alexander von Humboldt Foundation.

%% file: PRL_Btautau.bbl
\begin{thebibliography}{14}
\bibitem[{cry()}]{crystalnote}
\bibinfo{note}{The Crystal Ball function is defined to be \\ $X(m) = \left\{
  \begin{array}{l} (1 + \alpha (m-m_{0})/n \sigma - \alpha^{2}/n)^{-n} \exp
  \left[ - \frac{1}{2} \alpha^{2} \right] \\ \exp \left[ - (m-m_{0})^{2}/2
  \sigma^{2} \right] \end{array} \right\}$ \\ for $(m-m_{0}) < \sigma \alpha$
  (top) and $(m-m_{0}) \geq \sigma \alpha$ (bottom)}.
\bibitem[{arg()}]{argusnote}
\bibinfo{note}{The ARGUS function is defined to be \\ $A(m) = m \sqrt{1 -
  (m/m_{c}) ^{2}} \exp \left[ a \left( 1 - \left( m/m_{c} \right) ^{2} \right)
  \right]$}.
\bibitem[{cha()}]{chargenote}
\bibinfo{note}{Charge conjugate states are included implicitly throughout this
  paper.}
\end{thebibliography}
